\documentclass[aps,twocolumn,showpacs]{revtex4}

\usepackage{dcolumn}
\usepackage{graphicx}

\begin{document}

\title{Large well-relaxed models of vitreous silica, \\
coordination numbers and entropy}

\author{R. L. C. Vink}
\email{vink@phys.uu.nl}
\homepage{http://www.phys.uu.nl/~vink}
\affiliation{
        Institute for Theoretical Physics,
        Utrecht University,
        Leuvenlaan 4,
        3584 CE Utrecht, the Netherlands }

\author{G. T. Barkema}
\affiliation{
        Institute for Theoretical Physics,
        Utrecht University,
        Leuvenlaan 4,
        3584 CE Utrecht, the Netherlands }

\date{December 10, 2002}

\begin{abstract} 
A Monte Carlo method is presented for the simulation of vitreous silica.
Well-relaxed networks of vitreous silica are generated containing up to
300,000 atoms.  The resulting networks, quenched under the BKS potential,
display smaller bond-angle variations and lower defect concentrations,
as compared to networks generated with molecular dynamics. The total
correlation functions $T(r)$ of our networks are in excellent agreement
with neutron scattering data, provided that thermal effects and the
maximum inverse wavelength used in the experiment are included in
the comparison. A procedure commonly used in experiments to obtain
coordination numbers from scattering data is to fit peaks in $rT(r)$ with
a gaussian.  We show that this procedure can easily produce incorrect
results. Finally, we estimate the configurational entropy of vitreous
silica.
\end{abstract}



\maketitle

\section{Introduction}

If liquid silica (SiO$_2$) is cooled below its melting temperature,
it usually does not crystallize, but stays a supercooled liquid
for an extended period of time. Upon further cooling, to below a
certain temperature known as the glass transition temperature $T_g$,
it becomes highly viscous and shows many properties of ordinary solids,
but stays disordered and does not evolve to the thermodynamically stable
crystalline phase, at least on earthly time scales. The study of silica
in its disordered phase (vitreous silica) is hampered by a lack of
knowledge on its microscopic structure. Experimental techniques alone,
such as neutron and X-ray scattering, which can uniquely resolve the
structure of a crystal, are less successful in this case because of the
lack of a repeating unit cell. Typically, only information averaged over
many atoms, such as the radial distribution function, can be obtained
from experiments.

An alternative method to study vitreous silica is through computer
simulations. The usual computational approach to generate vitreous silica
structures is to simulate a quench from the melt within the framework of
molecular dynamics (MD). This resembles closely the experimental process
used to prepare vitreous silica, except that the typical computational
cooling rates are about ten orders of magnitude faster than experimental
ones. In agreement with experiment, these MD simulations result in
structures of vitreous silica in which almost all silicon atoms are bonded
to four oxygen atoms and almost all oxygen atoms are bonded to two silicon
atoms, without any sign of long-range order~\cite{vollmayr1996a}. This
in turn is in agreement with the continuous random network (CRN) model,
proposed by Zachariasen~\cite{zachariasen1932a}.

The high cooling rates used in MD simulations result in structures with a
strain higher than observed in experiments. This shows up, for instance,
as an anomalously high density of coordination defects and a larger spread
in bond lengths and bond angles. In this paper we present an alternative
approach to generate models of vitreous silica. In contrast to MD, we do
not attempt to simulate the details of the dynamics of the melt-quenching
process. Instead, we use a Monte Carlo (MC) scheme with an artificial
dynamics consisting of {\em bond transpositions}. Our scheme is similar in
spirit to the one used by Tu {\it et al.}~in their study of silica and
silica sub-oxides~\cite{tu1998a,tu2000a}, which in turn is based on the
algorithm of Wooten, Winer and Weaire (WWW) for the generation of
four-fold coordinated CRNs~\cite{wooten1985a,djordjevic1995a}.

The outline of this paper is as follows. We first describe the MC
scheme used by Tu {\it et al}. We then move on to describe a number
of optimizations to this algorithm. These optimizations allow us to
generate large and well-relaxed silica networks containing up to 300,000
atoms. The properties of these networks are discussed and compared to
MD-prepared networks and experiment in Section~\ref{sec:res}. We end
with a number of conclusions in Section~\ref{sec:con}.

\section{Bond-switching algorithm for silica}

In the approach followed by Tu and co-workers, a vitreous silica network
consists of $N_s$ silicon atoms and $N_o =2 N_s$ oxygen atoms. The total
number of atoms in the network thus equals $N=N_s+N_o$. At all times, an
explicit list of bonds is maintained which uniquely determines the
connectivity or topology of the network. In the list of bonds, each
silicon atom is bonded to four oxygen atoms, and each oxygen atom is
bonded to two silicon atoms. The total number of bonds in the list thus
equals $4N_s$.

To calculate energy and forces the following potential is used:
\begin{eqnarray}\label{eq:tt}
  E_{TT} &=& \frac{1}{2} \sum_i k_b (b_i - b_0)^2 \nonumber \\ 
  &+& \frac{1}{2} \sum_{i,j} k_\theta (\cos\theta_{ij} - \cos\theta_0)^2,
\end{eqnarray}
where the first summation runs over the list of bonds, and the second
summation over all pairs of bonds that share one atom; $b_i$ represents
the length of the $i$-th bond, and $\theta_{ij}$ is the angle between
bonds $i$ and $j$ (which share one atom). The potential parameters
for the two-body \mbox{Si-O} interaction are given by $k_b=27.0$~eV
and $b_0=1.60$~\AA. For the three-body \mbox{O-Si-O} interaction common
values are $k_\theta=4.32$~eV and $\cos\theta_0=-1/3$; for \mbox{Si-O-Si}
interactions these values are $k_\theta=0.75$~eV and $\cos\theta_0=-1$.
Since the number of bonds in the network grows linearly with the number
of atoms it contains, each energy evaluation is an $O(N)$ operation.

\begin{figure*}
\begin{center}
\includegraphics[width=12cm]{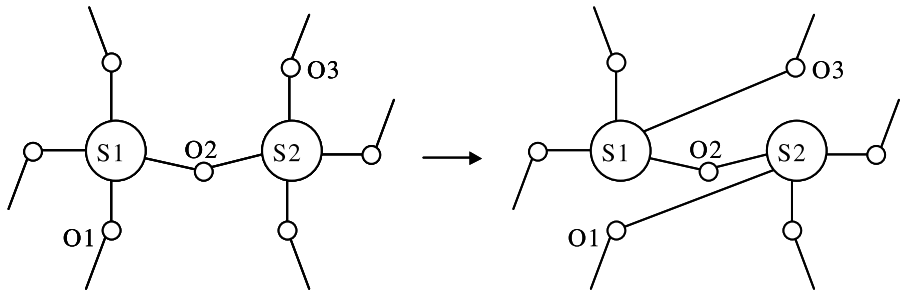}
\caption{\label{fig:wwwsilica}Bond transposition used in the generation of
vitreous silica structures. Two silicon atoms $\{S1,S2\}$ and three oxygen
atoms $\{O1,O2,O3\}$ are selected following the geometry shown left. Next,
bonds $S1-O1$ and $S2-O3$ are broken and two new bonds $S1-O3$ and $S2-O1$
are created to obtain the geometry shown right. After the bond transposition,
the geometric coordinates of the atoms are relaxed (not shown).}
\end{center}
\end{figure*}

The network is evolved by making explicit changes in the list of bonds,
each time followed by a local energy minimization with respect to the
geometric coordinates of all atoms (using for example the method of
steepest descent). The changes in the list of bonds consist of bond
transpositions as illustrated in Fig.~\ref{fig:wwwsilica}. Each bond
transposition is accepted with the Metropolis acceptance
probability~\cite{metropolis1953a}:
\begin{equation}\label{eq:metropolis}
  P_a = \min \left[1, \exp \left( \frac{E_b-E_f}{k_B T} \right)
\right],
\end{equation}
where $k_B$ is the Boltzmann constant, $T$ is the temperature, and $E_b$
and $E_f$ are the (minimized) energies of the network before and after
the proposed bond transposition.

\section{Optimized bond-switching algorithm for silica}
\label{sec:opt}

We have optimized the bond-switching algorithm of Tu {\it et al.}~for the
generation of large and well relaxed silica networks. These optimizations
are similar in spirit to the optimizations used in the scalable WWW
algorithm~\cite{barkema2000a,vink2001c}.

The first optimization is aimed at reducing the CPU time spent on rejected
bond transpositions. After a bond transposition in the original algorithm,
the energy of the network is always completely minimized. After the
minimization, the bond transposition is either accepted or rejected
based on the Metropolis probability.

In contrast, we reformulate the Metropolis algorithm. We first determine a
threshold energy $E_t= E - k_B T \ln(1-r)$, where $r$ is a random number
uniformly drawn from the interval $[0,1\rangle$. We then proceed with the
minimization procedure. During minimization, the converged energy is
continuously estimated. Assuming that the total energy is harmonic around
the minimum with curvature $c$, the energy $E(\vec{R}_0)$ at the (unknown)
minimum $\vec{R}_0$ can be estimated from a nearby configuration $\vec{R}$
as $E( \vec{R}_0) \approx E(\vec{R})  - |\vec{F}|^2/c$, with $|\vec{F}|$
the magnitude of the total force at $\vec{R}$. Minimization is aborted as
soon as it becomes clear that the threshold energy will not be reached and
the bond transposition will be rejected. This leads to a large reduction
in the number of force evaluations associated with rejected bond
transpositions.

The second optimization is aimed at exploiting the local nature of the
bond transposition depicted in Fig.~\ref{fig:wwwsilica}. Immediately after
a bond transposition, only a small cluster of atoms in the network will
experience a significant force. This cluster consists of the atoms
directly involved in the bond transposition, marked $\{S1,S2,O1,O2,O3\}$
in Fig.~\ref{fig:wwwsilica}, and of nearby atoms (typically extending to
the fourth neighbor shell around these five transposition atoms). The
number of atoms in such a cluster is about 400. It therefore suffices to
calculate the force {\it locally} (only for the 400 or so atoms inside the
cluster) rather than {\it globally} (for all the atoms in the network).
Calculating the force on a cluster of atoms is an $O(1)$ operation, which
means that it is independent of the total number of atoms in the network.
Local force calculations are therefore much cheaper than global $O(N)$
force calculations.

Already after a converged local minimization, it is often clear that
the threshold energy will not be reached, and the bond transposition
can be rejected. If this is not the case, a few global minimization
steps are required additionally, usually resulting in an accepted
bond transposition. The combination of these two improvements reduces
the computational effort from $O(N)$ operations per attempted bond
transposition to $O(1)$ operations per attempted bond transposition,
plus a few $O(N)$ operations per accepted bond transposition. Since
common acceptance probabilities are well below 1\%, the speed-up for
large systems is about two orders of magnitude.

\section{Initial configurations}

The optimized bond-switching algorithm for silica requires an initial
network, which should be free of coordination defects and crystalline
regions. As noted in Ref.~\onlinecite{thorpe1986a}, four-fold coordinated
CRNs decorated with one oxygen atom on each bond already provide
structures that compare reasonably well to vitreous silica. As initial
network we therefore use a periodic, four-fold coordinated CRN generated
as described in Refs.~\cite{barkema2000a,vink2001c}. This CRN serves as
the silicon backbone. Next, one oxygen atom is placed on the center of
each \mbox{Si-Si} bond to obtain a properly coordinated silica network.
The simulation volume is then scaled to obtain the experimental density of
silica $\rho = 2.20$~g~cm$^{-3}$~\cite{mazurin1983a}.

\section{BKS sampling}
\label{sec:bks}

A computationally more expensive, but also more realistic potential for
silica is the BKS potential~\cite{beest1990a}:
\begin{equation}
  E_{BKS} = \sum_{i<j} \left( \frac{ q_i q_j }{ r_{ij} } +
        A_{ij} \exp \left( -B_{ij} r_{ij} \right) -
        \frac{ C_{ij} }{ r_{ij}^6 } \right),
\end{equation}
where the sum runs over all pairs of atoms in the system, $r_{ij}$ is the
distance between atoms $i$ and $j$, and $q_i$ is the charge of atom $i$.  
The atomic charges and the values of the potential parameters $A_{ij}$,
$B_{ij}$ and $C_{ij}$ are also listed in Ref.~\onlinecite{beest1990a}. We
use the fast multipole code developed at Duke
University~\cite{rankin1995a} to compute the Coulomb and van der Waals
interactions. The exponential term is truncated at $5.50$~\AA\ and then
shifted, as described in Ref.~\onlinecite{vollmayr1996a}.

The bond-switching algorithm described above generates networks that are
local energy minima according to the potential of Tu {\it et al.} The
algorithm can be changed such that these networks will be sampled from
the Boltzmann distribution according to the BKS potential at temperature
$T$.  To this end, the acceptance probability of Eq.~(\ref{eq:metropolis})
is replaced by:
\begin{eqnarray}
  P &=& \min \left[1,
  \exp \left( -\frac{ \Delta E_{TT} }{ k_B T' } \right) \right] \nonumber \\
  &&\times \min \left[1,
	\exp \left( \frac{ \Delta E_{TT} }{ k_B T' } -
   \frac{ \Delta E_{BKS} }{ k_B T } \right) \right],
\end{eqnarray}
where $\Delta E_{TT}$ and $\Delta E_{BKS}$ are the energy differences
between the networks before and after the bond transposition, calculated
with the potential of Tu {\it et al.}~and the BKS potential, respectively.
The algorithm satisfies detailed balance, also when the fictitious
temperatures $T$ and $T'$ are different. In the expression above, the
first factor biases moves according to the potential of Tu {\it et al.}~at
temperature $T'$, after which the second factor corrects for the
difference between this potential at temperature $T'$ and the BKS
potential at temperature $T$.

With the above probability, bond transpositions that are accepted with the
potential of Tu {\it et al.}~undergo an additional accept/reject decision
based on the BKS potential. By increasing the temperature $T'$ we can
increase the number of moves that will be accepted under the potential of
Tu {\it et al}. In the limit $T' \rightarrow \infty$, all such moves are
accepted. In this limit, the accept/reject decision is based solely on the
BKS potential and $T$. While correct, this situation is undesirable
because BKS energy evaluations are expensive. In practice, a lower $T'$ is
used and the temperatures $\{T,T'\}$ are tuned to maximize the decrease of
the BKS energy of the network as a function of simulation time.

\section{Results}
\label{sec:res}

We have used the optimized bond-switching algorithm described in
Section~\ref{sec:opt} to generate vitreous silica networks containing
3000 atoms (`3k'), 60,000 atoms (`60k') and 300,000 atoms (`300k'). For
the 3000-atom and the 60,000-atom networks the bond-switching algorithm
was used in conjunction with BKS sampling at temperatures $k_B T=0.30$~eV
and $k_B T'=0.70$~eV (see Section~\ref{sec:bks}). For the 300,000-atom
network, BKS sampling proved to be too computationally demanding and
was not implemented; for this network the Metropolis probability of
Eq.~(\ref{eq:metropolis}) was used with $k_B T=0.15$~eV and the 
potential of Tu {\it et al}.

All three networks were evolved with approximately 10 attempted bond
transpositions per atom from their starting configurations. The energy
of the resulting networks was then minimized in a single quench using
the method of steepest descent and the BKS potential without volume
optimization. We observed that if volume optimization were used with the
BKS potential, the density of the networks becomes unphysically high,
typically in the range 2.27-2.37~g~cm$^{-3}$. Similar high densities are
also reported in a number of MD studies~\cite{vollmayr1996a,horbach1999a}.

\begin{figure}
\begin{center}
\includegraphics[width=9cm]{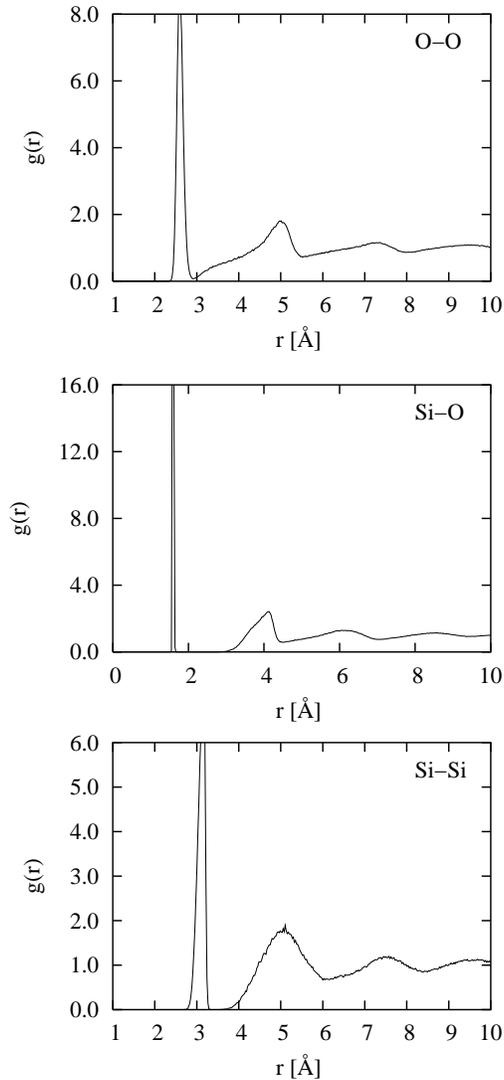}
\caption{\label{fig:rdf60k}Species-dependent density-normalized radial 
distribution functions $g_{\alpha\beta}(r)$ for network `60k' after 
relaxation at zero temperature with the BKS potential~\cite{beest1990a}.}
\end{center}
\end{figure}

In Fig.~\ref{fig:rdf60k} we show the radial distribution functions (RDFs)
$g_{SiSi}(r)$, $g_{SiO}(r)$, and $g_{OO}(r)$ for model `60k'. The RDFs are
density normalized such that \mbox{ $\lim_{r\rightarrow\infty}
g_{\alpha\beta}(r)=1$}. The RDFs for models `3k' and `300k' are similar
(not shown).

The Si-O RDF was used to extract a cutoff for the Si-O bond length at
$r_c=1.80$~\AA, where $r_c$ is taken at the minimum between the first and
the second peak. Based on this $r_c$ we find structural properties as
reported in Table~\ref{tab:comp}. For comparison, we also report results
obtained with MD using the same BKS potential. Compared to MD, we observe
that our networks are better relaxed, as is evident from the smaller
variations in the O-Si-O and Si-O-Si bond angles. Moreover, our networks
are nearly free of coordination defects.

\begin{table}
\caption{\label{tab:comp}Structural and energetic properties of our
vitreous silica networks at zero temperature, after a local energy
minimization under the BKS potential at the experimental density. For
comparison we also report properties of silica networks prepared by a
quench to zero temperature under MD~\cite{vollmayr1996a} using the same
potential. The number of atoms in each structure is given by $N$; $E$ is
the BKS energy per silicon atom in eV; $\rho$ is the density in
g~cm$^{-3}$; the Si-O bond length and its rms variation are given in \AA;
the mean bond angles and their variations (rms and FWHM) are given in
degrees; $z_4$ and $z_2$ are the percentages of perfectly coordinated
silicon and oxygen atoms, respectively, based on a \mbox{Si-O} bond-length
cut-off at 1.80~\AA.}
\vspace{2mm}
\begin{ruledtabular}
\begin{tabular}{l|cccc}
        & `3k'   & `60k'  & `300k'  & MD \\ \hline
$N$     & 3000   & 60,000 & 300,000 & 1002 \\
$E$     & -58.12 & -58.10 & -58.09  & -- \\
$\rho$  & 2.20\footnotemark[1] & 2.20\footnotemark[1] & 2.20\footnotemark[1] & 2.27-2.37 \\ \hline
Si-O    &        &        &         &           \\
mean    & 1.606  & 1.608  & 1.606   & 1.595\footnotemark[2]     \\
rms     & 0.010  & 0.011  & 0.011   & -- \\ \hline
O-Si-O  &        &        &         &           \\
mean    & 109.44 & 109.43 & 109.43  & 108.3     \\
rms     & 3.95   & 4.59   & 4.32    & -- \\
FWHM    & 8.3    & 9.3    & 9.8     & 12.8      \\ \hline
Si-O-Si &        &        &         &           \\
mean    & 153.89 & 153.57 & 153.00  & 152       \\
rms     & 11.75  & 11.72  & 11.94   & --  \\
FWHM    & 34     & 33.3   & 34.5    & 35.7      \\ \hline 
$z_4$   & 100\%  & 100\%  & 99.997\%  & 99.8\%     \\
$z_2$   & 100\%  & 100\%  & 99.998\%  & 99.8\%     \\  
\end{tabular}
\end{ruledtabular}
\vspace{2mm}
\footnotetext[1]{Fixed to the experimental density $\rho =
2.20$~g~cm$^{-3}$~\cite{mazurin1983a}.}
\footnotetext[2]{Location of the first peak in the \mbox{Si-O} RDF.}
\end{table}

A more stringent test is to compare the properties of our networks to
experimental data. The crucial quantity to consider is the total
correlation function $T(r)$ as obtained in X-ray or neutron scattering
experiments. In these experiments, the quantity that can be measured
directly is the total interference function $Qi(Q)$; the correlation
function $T(r)$ is obtained via a Fourier transform of $Qi(Q)$. In
practice, the resolution in $T(r)$ is determined by the maximum attainable
inverse wavelength $Q_{max}$ in the experiment. As discussed in
Ref.~\onlinecite{wright1994a}, peaks in $T(r)$ are significantly broadened
because $Q_{max}$ is not infinite in real experiments. In addition to
Fourier broadening, there is thermal broadening because experiments are
typically carried out at room temperature. For a meaningful comparison it
is essential to take these factors into account.

To capture the effect of thermal broadening, all three networks
are thermalized over approximately 1~ps with MD using the BKS
potential in the NVE-ensemble at $T\approx 300$~K. The structural
properties of the thermalized networks are summarized and compared
to experiment in Table~\ref{tab:thermal}. We observe good agreement
on the \mbox{Si-O} bond length, the variation in the bond length and
the mean bond angles. Compared to experiment, our networks slightly
overestimate the variation in the \mbox{O-Si-O} bond angle. This
may indicate that experimental vitreous silica is more relaxed
than our networks. Experimental estimates for the variation in the
\mbox{Si-O-Si} bond angle range from 12.8~degrees~\cite{grandinetti2002a}
to 36~degrees~\cite{mozzi1969a}. Compared to these data our networks
coincide with the higher value. The coordination numbers reported in
Table~\ref{tab:thermal} indicate that the vast majority of atoms is
still properly coordinated after thermalization.

\begin{table}
\caption{\label{tab:thermal}Structural properties of networks `3k', `60k'
and `300k' after thermalization at 300~K, based on a \mbox{Si-O}
bond-length cut-off at $1.80$ \AA.  Shown are the mean \mbox{Si-O} bond
length with rms variation in \AA; the mean bond angles with rms variation
and FWHM of the corresponding distribution in degrees; and the mean
coordination number $z$ of silicon atoms. For comparison we also report
values obtained in experiments.}
\vspace{2mm}
\begin{ruledtabular}
\begin{tabular}{l|cccc}
        & `3k'   & `60k'  & `300k'  & Experiment                \\ \hline
Si-O    &        &        &         &                           \\
mean    & 1.610  & 1.614  & 1.612   & 1.609~\cite{grimley1990a}  \\
rms     & 0.042  & 0.042  & 0.041   & 0.047~\cite{grimley1990a}  \\ \hline
O-Si-O  &        &        &         &                           \\
mean    & 109.38 & 109.37 & 109.37  & $109.7\pm0.6$~\cite{grimley1990a} \\
rms     & 5.79   & 6.27   & 6.31    & 4.5~\cite{grimley1990a}   \\
FWHM    & 13.1   & 14.4   & 14.5    & --                        \\ \hline
Si-O-Si &        &        &         &                           \\
mean    & 153.01 & 152.74 & 152.20  & 148~\cite{grandinetti2002a};152~\cite{coombs1985a} \\
rms     & 12.22  & 12.31  & 12.60   & 7.5~\cite{neuefeind1996a} \\
FWHM    & 35     & 34.4   & 36.2    & 12.8~\cite{grandinetti2002a};
                                      26~\cite{pettifer1988a};
                                      36~\cite{mozzi1969a}      \\ \hline
$z$     & 4.000  & 4.000  & 4.000   & -- \\ 
\end{tabular}
\end{ruledtabular}
\end{table}

Next, Fourier broadening is taken into account. In the absence of Fourier
broadening, the total correlation function of a network is simply a
weighted sum of the RDFs:
\begin{eqnarray}\label{eq:tr}
  T(r) &=&  \frac{r}{w_{SiSi} + w_{SiO} + w_{OO}} 
  	\left[ w_{SiSi} \times g_{SiSi}(r) \right. \nonumber \\[3mm] 
  &&+ \left. w_{SiO} \times g_{SiO}(r) + 
	w_{OO} \times g_{OO}(r) \right],
\end{eqnarray}
in which each RDF is normalized such that $\lim_{r \rightarrow \infty}
T(r)=r$. We focus on neutron scattering, in which case the weights
are given by: $w_{SiSi}=0.1722$, $w_{SiO}=2\times0.4817$ and $w_{OO}
= 4\times0.3370$, see Ref.~\onlinecite{wright1994a}. These weights
include the neutron scattering lengths of silicon and oxygen atoms;
the factors of 2 and 4 account for the chemical composition of silica
(two oxygen atoms for every silicon atom).

To obtain the Fourier-broadened correlation function $T_b(r)$, the
correlation function of Eq.~(\ref{eq:tr}) is transformed to obtain
the interference function $Qi(Q)$:
\begin{equation}
  Qi(Q) = \int_{r=0}^{r=r_{max}} \left[ T(r)-r \right] \sin(Q r) dr, 
\end{equation}
where $r_{max}$ is half the edge of the cubic simulation volume of the
model. Next, we transform the interference function back to 
obtain $T_b(r)$:
\begin{equation}
  T_b(r) = r +
        \frac{2}{\pi} \int_{Q=0}^{Q=Q_{max}} M(Q) Qi(Q) \sin(Q r) dQ,
\end{equation}
using the experimentally relevant $Q_{max}$. The function $M(Q)$ is an
(optional) modification function commonly used to reduce Fourier
artifacts. A popular choice is the Lorch modification
function~\cite{lorch1969a} given by:
\begin{equation}
  M(Q) = \frac{ Q_{max} }{ \pi Q }
        \sin \left( \frac{ \pi Q}{Q_{max} } \right).
\end{equation}

\begin{figure}
\begin{center}
\includegraphics[width=9cm]{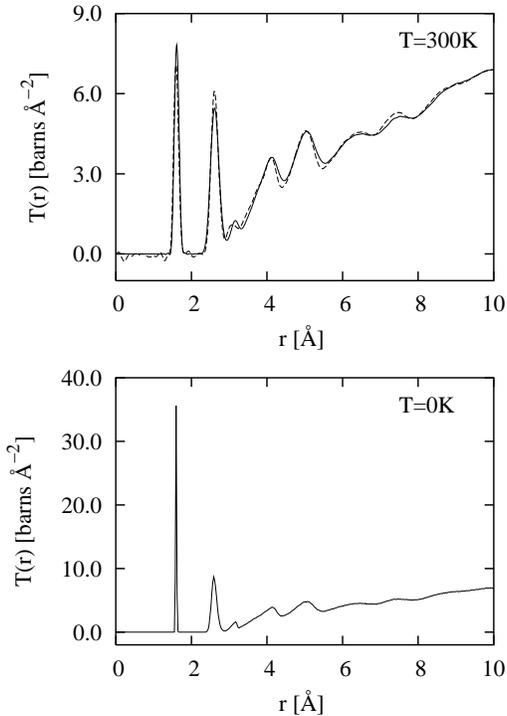}
\caption{\label{fig:tr}Total neutron correlation function $T(r)$ of
vitreous silica. The top frame shows the Fourier-broadened correlation
function of thermalized network `60k' (solid) compared to the experimental
result of Wright and co-workers (dashed) taken from
Ref.~\onlinecite{wright1991a}. The lower frame shows the correlation
function of the same network at zero temperature without Fourier
broadening.}
\end{center}
\end{figure}

The neutron $T(r)$ of vitreous silica was accurately measured by Grimley,
Wright and co-workers in 1991~\cite{grimley1990a,wright1991a} using
high-energy neutrons with $Q_{max}=45.2$ \AA$^{-1}$. For each of the three
thermalized networks we determined the Fourier-broadened correlation
function $T_b(r)$ using the same value for $Q_{max}$ and the Lorch
modification function. The experimental correlation function and the
correlation function of the thermalized '60k' network are compared to each
other in Fig.~\ref{fig:tr}. We observe excellent agreement. As a more
quantitative measure of the agreement between model and experiment we
consider the $R_\chi$ factor~\cite{wright1994a} defined as:
\begin{equation}
  R_\chi^2 = \frac{ \int \left[ T_{exp}(r)-T_{model}(r) \right]^2 dr }
			{ \int T_{exp}^2(r) dr }.
\end{equation}
The discrepancy between the experimental $T(r)$ and $T_b(r)$ of the
thermalized `60k' network, over the range $1.0 \leq r \leq 10.0$ \AA,
equals $R_\chi=5.1\%$. For the thermalized networks `3k' and `300k' we
obtain $R_\chi$ factors of 4.7\% and 4.9\%, respectively, over the same
range. 

\begin{figure}
\begin{center}
\includegraphics[width=9cm]{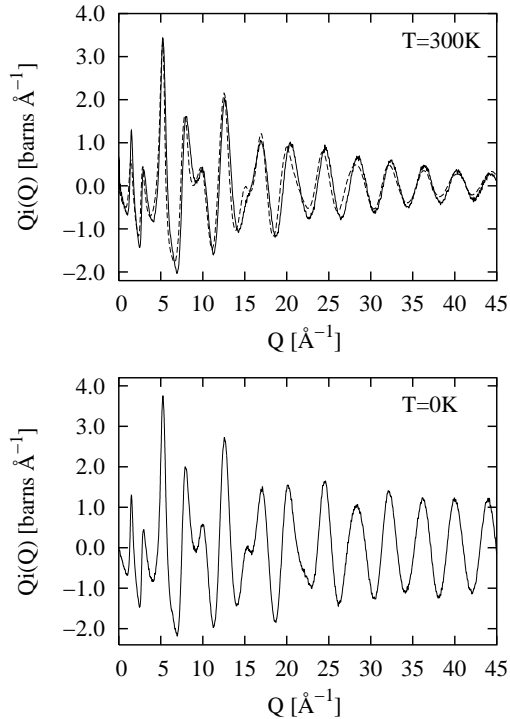}
\caption{\label{fig:sq}Total neutron interference function $Qi(Q)$ of
vitreous silica. The top frame shows $Qi(Q)$ of thermalized network
`60k' (solid) compared to experimental data (dashed) of Grimley and
co-workers~\cite{grimley1990a}. The lower frame shows the interference
function of network `60k' at zero temperature.}
\end{center}
\end{figure}

In the top frame of Fig.~\ref{fig:sq} we compare the total interference
function of the thermalized `60k' network to experiment. There is excellent
agreement on the overall peak positions and also on the damping of the
interference function for large $Q$. We observe some discrepancy at
$Q\approx15$~\AA$^{-1}$, where the small peak visible in the experimental
data is not reproduced by our network. The discrepancy seems to be
temperature-related: the lower frame of Fig.~\ref{fig:sq} shows that
the network at zero temperature does reproduce the peak correctly.

We observe in Fig.~\ref{fig:tr} that the total correlation function is
significantly broadened by Fourier and thermal effects, in particular
the first peak at $r \approx 1.6$ \AA. We have quantified the broadening
of this peak by fitting it to the form:
\begin{equation}\label{eq:pw}
  P(r) = \frac{1}{\sqrt{2\pi}} \frac{\eta}{r \sigma} 
	\exp \left[ -\frac{ (r-r_0)^2}{2\sigma^2} \right],
\end{equation}
with fit parameters $\{r_0,\sigma,\eta\}$. This follows the same procedure
used in Ref.~\onlinecite{grimley1990a} to obtain bond lengths and
coordination numbers from experimentally obtained correlation functions.
The parameters $r_0$ and $\sigma$ provide estimates for the mean
\mbox{Si-O} bond length and its variation. The parameter $\eta$ can be
used to extract the mean coordination number $z$ of silicon atoms:
\begin{equation}
 z = 4 \pi \eta \rho_O 
	\frac{ w_{SiSi} + w_{SiO} + w_{OO} }{ w_{SiO} },
\end{equation}
with $\rho_O$ the number density of oxygen atoms, $w_{\alpha\beta}$ the
neutron weights and $T(r)$ normalized such that $\lim_{r \rightarrow
\infty} T(r)=r$. 

The effects of Fourier and thermal broadening on the parameters
$\{r_0,\sigma,z\}$ are reported in Table~\ref{tab:ph}. As expected, the
broadening is significant: the combined effect of Fourier and thermal
broadening will boost $\sigma$ by a factor of over five. 

It is also interesting to compare the silicon coordination number obtained
directly from the atomic coordinates of a network, to that obtained from a
fit to the correlation function of that network. To this end the data of
the networks at 300~K without Fourier broadening in Table~\ref{tab:ph} are
compared to Table~\ref{tab:thermal}. The fitting procedure accurately
predicts the \mbox{Si-O} bond length and its variation $\sigma$, but not
the coordination number which we know to be 4.000. This result is
important because it shows that in the ideal situation free of Fourier
broadening, fits based on Eq.~(\ref{eq:pw}) will systematically
underestimate the coordination number. The cause of this underestimation
is due to the non-gaussian nature of the peak in $rT(r)$: fitting with a
more elaborate function, for instance with a sum of two gaussians, yields
the correct coordination number. The table also shows that fits with
Eq.~(\ref{eq:pw}) applied to Fourier-broadened data tend to overestimate
the coordination numbers.

\begin{table}
\caption{\label{tab:ph}The effect of Fourier broadening (FB) and thermal
broadening on the optimal parameters $\{r_0,\sigma,z\}$ obtained from a
fit of Eq.~(\ref{eq:pw}) to the first peak in $T(r)$ of our networks.  
Network `3k' proved to be too small to accurately sample the first peak
and was not used in this comparison. The parameters $r_0$ and $\sigma$ are 
given in \AA.}
\vspace{2mm}
\begin{ruledtabular}
\begin{tabular}{l|ccc}
          &          & `60k'           & `300k'  \\ \hline
$T=0$~K   & $r_0$    & 1.607           & 1.601   \\
no FB     & $\sigma$ & 0.011           & 0.011   \\
          & $z$      & $3.962\pm0.011$ & $3.980\pm0.004$ \\ \hline
$T=0$~K   & $r_0$    & 1.609           & 1.607   \\
with FB   & $\sigma$ & 0.048           & 0.048   \\
          & $z$      & $4.155\pm0.052$ & $4.161\pm0.053$ \\ \hline
$T=300$~K & $r_0$    & 1.611           & 1.610   \\
no FB     & $\sigma$ & 0.041           & 0.040   \\
          & $z$      & $3.977\pm0.013$ & $3.978\pm0.011$ \\ \hline
$T=300$~K & $r_0$    & 1.614           & 1.612   \\
with FB   & $\sigma$ & 0.060           & 0.060   \\
          & $z$      & $4.046\pm0.019$ & $4.048\pm0.019$ \\ 
\end{tabular}
\end{ruledtabular}
\vspace{2mm}
\end{table}

\section{Configurational entropy}

Recently, we developed a method to determine the configurational
entropy of a network~\cite{vink2002a}. The method requires only
the atomic coordinates and a list of bonds between particles, for a
single well-relaxed configuration. In the original paper, the method
was applied to a silica network consisting of 3000 atoms. Starting
with the atomic coordinates of this network, the list of bonds was
constructed using $r_c=1.80$~\AA\ for the \mbox{Si-O} bond-length cutoff.
Each \mbox{Si-O-Si} bridge was replaced by a single \mbox{Si-Si} bond. We
then applied our method to the resulting silicon backbone network to
obtain for the configurational entropy $s=0.88$~$k_B$ per silicon atom.

The accuracy of the method is sensitive to the size of the network. As
mentioned in the original paper, the configurational entropy of vitreous
silica reported above is most likely an underestimate due to the limited
size of that network. The large networks generated in this work allow us
to quantify these finite size effects. They also allow us to determine the
minimum size of a network that is required to accurately measure the
configurational entropy.

To this end, we use our method to determine the entropy of networks `3k',
`60k' and `300k' in two different ways: first, by using the entire
simulation volume of the network; secondly, by using only half the volume
of the simulation cell. The first procedure yields entropies of 0.83, 0.99
and 1.04~$k_B$ per silicon atom, respectively, while the second procedure
yields 0.78, 0.97 and 1.03~$k_B$ per silicon atom, respectively. This
illustrates that finite-size effects are significant for the smallest
network, but small for the large ones. Most of the difference in entropy
between the three networks is due to the varying degree of relaxation in
these networks.

\section{Conclusions}
\label{sec:con}

In summary, we have presented a Monte Carlo-based approach for the
generation of well-relaxed networks of vitreous silica, based on earlier
work of Tu {\it et al}. With this method, networks containing 3000, 60,000
and 300,000 atoms are generated.  Compared to networks generated with
molecular dynamics, our networks have smaller bond-angle variations and
are nearly defect-free, indicating they are better relaxed.

The total correlation function $T(r)$ of our networks is in excellent
agreement with neutron scattering experiments (with $R_\chi$ factors
around 5\%), provided that thermal effects and the maximum experimental
inverse wavelength $Q_{max}$ are included in the comparison. We also
determined the silicon coordination number of our networks by fitting a
gaussian to $rT(r)$, as is commonly done in experiments. We observe that
the results of this procedure are biased to lower numbers by the
non-gaussian nature of the peaks, and to higher numbers due to finite
value of $Q_{max}$. The configurational entropy of vitreous silica was
determined to be 0.99~$k_B$ and 1.04~$k_B$ per silicon atom, for the
networks containing 60,000 and 300,000 atoms, respectively.

Upon request, we will communicate the atomic coordinates of our networks.

\bibliographystyle{prsty}

\begin{thebibliography}{99}

\bibitem{vollmayr1996a} K. Vollmayr, W. Kob, and K. Binder, Phys. Rev. B
{\bf 54}, 15808 (1996).

\bibitem{zachariasen1932a} W. H. Zachariasen, J. Am. Chem. Soc. {\bf 54},
3841 (1932).

\bibitem{tu1998a} Yuhai Tu, G. Grinstein, and David Vanderbilt, Phys. Rev. 
Lett. {\bf 81}, 4899 (1998).

\bibitem{tu2000a} Yuhai Tu and J. Tersoff, Phys. Rev. Lett. {\bf 84}, 4393
(2000).

\bibitem{wooten1985a} F. Wooten, K. Winer, and D. Weaire, Phys. Rev. Lett.  
{\bf 54}, 1392 (1985).

\bibitem{djordjevic1995a} B. R. Djordjevi\'c, M. F. Thorpe, and F. Wooten,
Phys. Rev. B {\bf 52}, 5685 (1995).

\bibitem{metropolis1953a} N. Metropolis, A. W. Rosenbluth, M. N.  
Rosenbluth, A. H. Teller, and E. Teller, J. Chem. Phys. {\bf 21}, 1087
(1953).

\bibitem{barkema2000a} G. T. Barkema and N. Mousseau, Phys. Rev. B {\bf
62}, 4985 (2000).

\bibitem{vink2001c} R. L. C. Vink, G.T. Barkema, M.A. Stijnman, and R.H. 
Bisseling, Phys. Rev. B {\bf 64}, 245214 (2001).

\bibitem{thorpe1986a} M. F. Thorpe and S. W. de Leeuw, Phys. Rev. B {\bf
33}, 8490 (1986).

\bibitem{mazurin1983a} O. V. Mazurin, M. V. Strelsina, and T. P.
Shvaikovskaya, {\it Handbook of Glass Data} (Elsevier, Amsterdam, 1983),
Pt.~A.

\bibitem{beest1990a} B. W. H. van Beest, G. J. Kramer, and R. A. van
Santen, Phys. Rev. Lett. {\bf 64}, 1955 (1990).

\bibitem{rankin1995a} W. T. Rankin and J. A. Board, Technical Report
95-002, Department of Electrical Engineering, Duke University (1995).

\bibitem{horbach1999a} J. Horbach and W. Kob, Phys. Rev. B {\bf 60}, 3169 
(1999).

\bibitem{wright1994a} A. C. Wright, J. Non-Cryst. Solids {\bf 179}, 84
(1994).

\bibitem{grandinetti2002a} P. Grandinetti, based on recent NMR
measurements, private communication (2002).

\bibitem{mozzi1969a} R. L. Mozzi and B. E. Warren, J. Appl. Crystallogr.
{\bf 2}, 164 (1969).

\bibitem{lorch1969a} E. A. Lorch, J. Phys. C2, 229 (1969).

\bibitem{grimley1990a} D. I. Grimley, A. C. Wright, and R. N. Sinclair, J.
Non-Cryst. Solids {\bf 119}, 49 (1990).

\bibitem{wright1991a} A. C. Wright, R. A. Hulme, D. I. Grimley, R. N.
Sinclair, S. W. Martin, D. L. Price, and F. L. Galeener, J. Non-Cryst.  
Solids {\bf 129}, 213 (1991).

\bibitem{vink2002a} R. L. C. Vink and G. T. Barkema, Phys. Rev. Lett. {\bf
89}, 76405 (2002).

\bibitem{coombs1985a} P. G. Coombs, J. F. De Natale, P. J. Hood, E. K.
McElfresh, R. S. Wortman, and J. F. Schackelford, Philos. Mag. {\bf 51},
L39 (1985).

\bibitem{neuefeind1996a} J. Neuefeind and K. D. Liss, Ber.
Bunsengesellschaft {\bf 100}, 1341 (1996).

\bibitem{pettifer1988a} R. F. Pettifer, R. Dupree, I. Farnan, and U.
Sternberg, J. Non-Cryst. Solids {\bf 106}, 408 (1988).

\end{thebibliography}

\end{document}